\documentclass[aps,prb,reprint,twocolumn,superscriptaddress,notitlepage,showpacs]{revtex4}

\usepackage{amssymb,amsfonts,amsmath} 

\usepackage[pdftex]{graphicx} 
\usepackage{epstopdf}

\usepackage[utf8]{inputenc}
\usepackage[T1]{fontenc}
\usepackage{color}

\newcommand{\aver}[1]{\left\langle #1 \right\rangle}
\newcommand{\ket}[1]{\left | #1 \right\rangle}
\newcommand{\bra}[1]{\left\langle #1 \right |}

\begin{document}


\title{Tunable critical supercurrent and spin-asymmetric Josephson effect in superlattices}



\author{Juha~M.\ Kreula}
\affiliation{Clarendon Laboratory, University of Oxford, Parks Road, Oxford OX1 3PU, United Kingdom}
\affiliation{COMP Centre of Excellence, Department of Applied Physics, Aalto University, FI-00076 Aalto, Finland}

\author{Miikka~O.~J.\ Heikkinen} 
\affiliation{COMP Centre of Excellence, Department of Applied Physics, Aalto University, FI-00076 Aalto, Finland} 

\author{Francesco Massel} 
\affiliation{Department of Mathematics and Statistics, University of Helsinki, FI-00014 Helsinki, Finland}

\author{P\"aivi T\"orm\"a}
\altaffiliation{Electronic address: paivi.torma@aalto.fi}
\affiliation{COMP Centre of Excellence, Department of Applied Physics, Aalto University, FI-00076 Aalto, Finland} 




\begin{abstract}
Combining the Josephson effect with magnetism, or spin dependence in general, creates novel physical phenomena. The spin-asymmetric Josephson effect is a predicted phenomenon where a spin-dependent potential applied across a Josephson junction induces a spin-polarized Josephson current. Here, we propose an approach to observe 
the spin-asymmetric Josephson effect with spin-dependent superlattices, realizable, e.g., in ultracold atomic gases. We show that observing this effect is feasible by studying numerically the quantum dynamics of the system in one dimension. Furthermore, we
show
that the enhancement, or tunability, of the critical supercurrent in ferromagnetic Josephson junctions [F.\ S.\ Bergeret, A.\ F.\ Volkov, and K.\ B.\ Efetov, Phys.\ Rev.\ Lett.\ {\bf{86}}, 3140 (2001)] can be explained by the spin-asymmetric Josephson effect. 
\end{abstract}


\maketitle


\section{Introduction}
The Josephson effect~\cite{Josephson1962} is a consequence of a 
macroscopic phase coherence in superfluid condensates.
The phenomenon has played a significant role in topics ranging from basic
research on superconductivity and superfluidity to applications in electronics \cite{Barone1982}. 
The Josephson effect in combination with magnetism,
or spin-dependence in the generic case, has yielded several new physical phenomena.
Examples include $\pi$ junctions~\cite{Ryazanov2001}, spin-triplet Cooper-pair current~\cite{Khaire2010},
and the enhancement, or tunability, of the critical current 
in superconductor-ferromagnet structures~\cite{Bergeret2001}.
Another, striking prediction concerning spin-dependent Josephson systems
is the existence of a spin-asymmetric Josephson effect in which the Cooper-paired spins display frequency-synchronized oscillations with spin-dependent amplitudes~\cite{sorin,Heikkinen2010}.
Traditionally, the Josephson effect has been understood as the 
coherent tunneling of bosons, either elementary or composite, as in the case
of Cooper pairs, with no significant difference between these two cases~\cite{deGennes1999}.
However, the spin-asymmetric Josephson effect~\cite{sorin,Heikkinen2010} shows that in fermionic condensates the composite nature of Cooper pairs is always important and manifests itself in a dramatic way as a spin-polarized Josephson current.

\begin{figure}[ht!]
\centerline{\includegraphics[width=1.0\columnwidth]{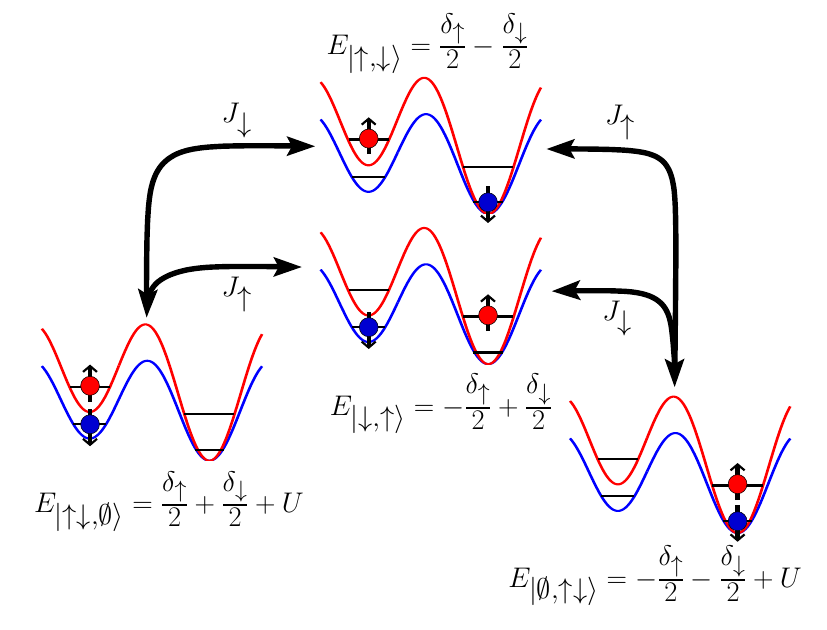}}
\caption{Origin of the spin-asymmetric Josephson effect illustrated in
a four-state model consisting of a
spin-dependent double well with two spins and on-site interaction $U$.
The initial state of the system is a superposition of the paired states, 
${|\psi(t=0)\rangle=\alpha |\uparrow \downarrow, \emptyset \rangle + \beta |\emptyset, \uparrow \downarrow \rangle}$.
The tunneling couplings $J_\uparrow$ and $J_\downarrow$ cause Josephson
oscillations in the system. The couplings
connect the paired states via intermediate states of broken pairs,
$| \uparrow, \downarrow \rangle$ and $| \downarrow, \uparrow \rangle$,
which are populated in the Josephson oscillations.
The spin-dependent potentials $\delta_\uparrow$ and $\delta_\downarrow$
create an energy difference between these intermediate states, 
resulting in different populations on these states, 
and thus the spin-asymmetric Josephson effect. 
}
\label{fig:basics1}
\end{figure}

In this work, we suggest an experimental arrangement to detect the yet-unobserved 
spin-asymmetric Josephson effect. The proposed setup is a spin-dependent superlattice, 
realizable, e.g., in ultracold atomic gas systems~\cite{stringarireview,blochreview}.
We simulate the quantum dynamics of the system in the case of
a one-dimensional (1D) superlattice, utilizing
the time-evolving block decimation (TEBD) method \cite{vidal2003,daley2004,vidal2004}.
Our results indicate that the observation of the spin-asymmetric Josephson effect 
is feasible with existing experimental techniques. 
Furthermore, we show that the spin-asymmetric Josephson effect elucidates the physical origin of the predicted enhancement~\cite{Bergeret2001}, in general tunability,
of the critical direct current (dc) in superconductor-ferromagnet structures. Detecting the spin-asymmetric Josephson effect would provide fundamental understanding of macroscopic quantum coherence and anticipate highly tunable Josephson devices.

In the spin-asymmetric Josephson effect the system of interest is conceptually
a Josephson junction where the two spin components of a Cooper pair
are subjected to different potentials $\delta_\uparrow$ and $\delta_\downarrow$, while the junction barrier 
is an insulator with no additional spin-dependence. 
The spin components oscillate at the same Josephson frequency 
but, suprisingly, with different amplitudes, yielding the Josephson current 
${I^J_{\sigma}(t) = I^C_{\sigma}(\delta_{-{\sigma}})\sin\left[ \left(\delta_\downarrow + \delta_\uparrow \right)t +\varphi \right]}$. 
Here, $\sigma=\uparrow,\downarrow$, $-\sigma$ denotes the opposite spin, 
and $\varphi$ is the initial phase difference. The asymmetric oscillations are  
in sharp contrast to the usual view
of composite boson tunneling and occur even though the pairing is of singlet-type, and no triplet-pairing is required. 
The effect is best understood in terms of a four state model 
of Fig.~\ref{fig:basics1} which demonstrates the dynamics of a single Cooper pair initially in 
a coherent superposition across the tunneling barrier~\cite{Heikkinen2010}. 
This model elucidates that the tunneling occurs through intermediate
states consisting of $\uparrow$ and $\downarrow$ spin components on different sides
of the junction. The salient point is that the Josephson current contains \textit{single-particle} 
interference terms occurring on these intermediate states, 
even at zero temperature and in the absence of initial quasiparticle excitations. Importantly, these interferences 
are different for the current of each component in the presence 
of spin-asymmetric potentials, leading to the spin-asymmetric Josephson effect.


\section{Ferromagnetic Josephson junctions and spin-asymmetric Josephson effect}
The spin-asymmetric potential is in close analogy to an 
exchange field in a ferromagnet. However, the spin-asymmetric Josephson 
effect is fundamentally different from the critical current reversal in 
superconductor-ferromagnet-superconductor (SFS) $\pi$-junctions~\cite{Ryazanov2001}
and from the spin-triplet supercurrent discovered
in multilayered ferromagnetic Josephson junctions~\cite{Khaire2010}.
In these phenomena
the barrier between the superconductors plays the key role~\cite{Golubov2004,Buzdin2005,Bergeret2005}, while in 
the spin-asymmetric Josephson effect 
the barrier can be just an insulator with no preference on spin, 
and the underlying physics results from the spin-dependent potentials instead.

\begin{figure}[ht!]
\centerline{\includegraphics[width=1.03\columnwidth]{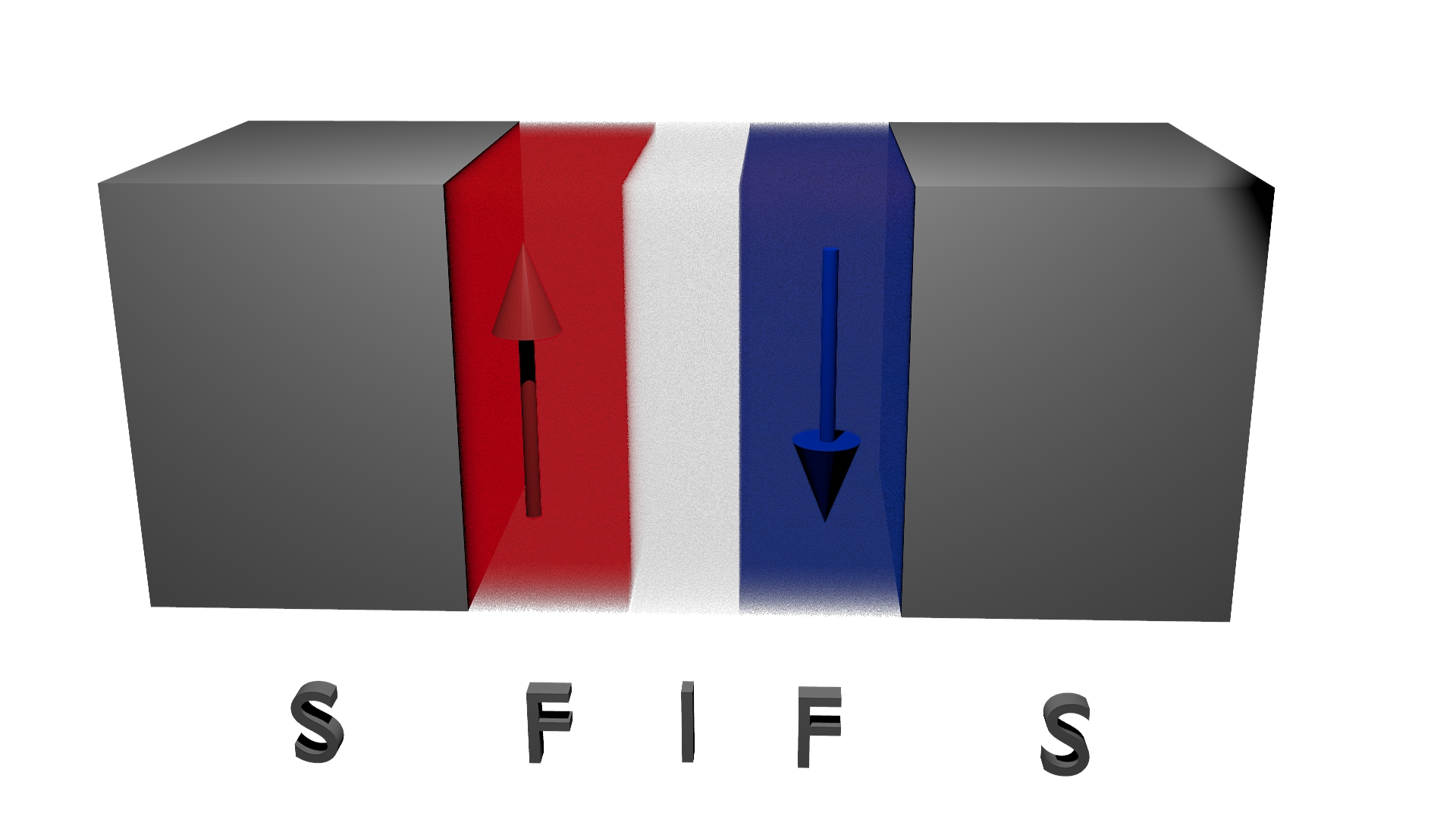}}
\caption{
SFIFS Josephson junction (S stands for superconductor, F for ferromagnet,
and I for insulator) with anti-parallel magnetizations in the F layers. 
When the SF bilayer can be considered a homogeneous
magnetic superconductor, the critical current in the junction
can be tuned by the exchange potential of the ferromagnet
in a scenario which is closely related to 
the spin-asymmetric Josephson effect.
}
\label{fig:basics2}
\end{figure}

Here we demonstrate that the spin-asymmetric Josephson effect is more closely related to a phenomenon which
can occur in an SFIFS (I stands for insulator) junction, 
see Fig.~\ref{fig:basics2}.
It has been predicted that, for antiparallel magnetizations in the F layers,
the critical dc Josephson current can be tuned by 
varying the magnetization~\cite{Bergeret2001,Golubov2002,Bruder2002}. 
This happens when the SF-bilayer structure can be considered
a uniform magnetic superconductor (note that a very different type of behavior is
anticipated for instance in long junctions~\cite{Hekking2004}). The assumption for the uniform magnetic superconductor holds when the thickness of the S layer is less than the superconducting coherence length, and the 
thickness of the F layer is less than the condensate penetration length to the ferromagnet~\cite{Bergeret2001}.
The uniform magnetization plays a role similar to the spin-asymmetric potential but
an essential difference is that the exchange field of 
the ferromagnet affects also the ground state of the system,
unlike in the spin-asymmetric Josephson effect.
However, we show that at zero temperature these two scenarios lead to the same outcome.

The derivation of the result follows closely the standard linear response description of the
Josephson effect, and we present it in detail in Appendix~\ref{app:connection}.
Here, we give only the result. 
We obtain the following form for the critical current of the $\uparrow$-component
in the SFIFS-junction with antiparallel magnetizations
\begin{align}\label{criticalSF}
I_\uparrow^C(h_R-h_L)\Big \vert_{\delta_{\uparrow/\downarrow}=0}&=2\sum_{kp}\Omega_{kp}\Omega_{-k,-p}u_k v_ku_p v_p \nonumber \\
&\times  \Bigg[ \frac{1}{E_0(k)+E_0(p)+h_R-h_L}\nonumber \\
&+\frac{1}{E_0(k)+E_0(p)-(h_R-h_L)}  \Bigg].
\end{align}
The critical current of the $\downarrow$-component has a similar expression.
In the above expression, $\Omega_{k,p}$ is the tunneling matrix element which couples the momentum states $|k\rangle$ and $|p\rangle$ on the left and right sides of the junction, respectively. Furthermore, $h_R$ is the magnetization on the right ferromagnetic superconductor (similarly for the left), $E_0(k)=\sqrt{\xi_k^2+\Delta^2}$, where 
$\xi_k$ is the single electron dispersion relative to the chemical potential, 
$\Delta$ denotes the BCS order parameter, and $u_k$ and $v_k$ are the 
Bogoliubov coefficients given by
\begin{align}
u_k&=\sqrt{\frac{1}{2}\left(1+\frac{\xi_k}{\sqrt{\xi_k^2+\Delta^2}}\right)},\\
v_k&=\sqrt{\frac{1}{2}\left(1-\frac{\xi_k}{\sqrt{\xi_k^2+\Delta^2}}\right)}.
\end{align}

In the case of the spin-asymmetric Josephson effect, the critical $\uparrow$-current reads
\begin{align}\label{criticalSAJE}
&I_\uparrow^C(-\delta_\downarrow)\Big \vert_{h_{L/R}=0}=2\sum_{kp}\Omega_{kp}\Omega_{-k,-p} u_k v_k u_p v_p\nonumber \\ &\times
\Bigg[ \frac{1}{E_0(k)+E_0(p) +\delta_\downarrow}+\frac{1}{E_0(k)+E_0(p)-\delta_\downarrow}\Bigg].
\end{align}

Comparing Eqs.~\eqref{criticalSF} and \eqref{criticalSAJE}, we conclude that the form of the critical current is the same in the magnetically tuned SFIFS junction and in the Josephson junction driven by spin-asymmetric potentials. 
Thus, the tunable dc supercurrent in SFIFS junctions 
can be considered the dc limit ($\delta_\downarrow + \delta_\uparrow = 0$) of the spin-asymmetric Josephson effect 
with the remaining degree of freedom,
$\delta_\uparrow = - \delta_\downarrow$, corresponding to antiparallel magnetization. 
We stress that the dc limit of the spin-asymmetric Josephson effect 
corresponds to the condition 
$\delta_\downarrow=-\delta_\uparrow$, 
not only to the case 
when both of the potentials are zero.

The origin of the tunable critical current can be explained in terms of Fig.~\ref{fig:basics1}. 
The energies of the intermediate, 
broken Cooper pair states depend on $|\delta_\uparrow|=|\delta_\downarrow|\equiv\delta$
[i.e.\ $I^C_\uparrow (\delta) = I^C_\downarrow (\delta)$],
which allows the tuning of the amplitude of 
the supercurrent by varying $\delta$
without changing the Josephson frequency from zero.
In Appendix~\ref{4state} we elaborate on how this argument, presented originally
for a momentum conserving tunneling coupling, is extended to a general (non-momentum conserving) 
tunneling coupling which appears in Eqs.~\eqref{criticalSF} and~\eqref{criticalSAJE} above.
Furthermore, we emphasize that the spin-asymmetric Josephson effect predicts the 
remarkable possibility of tuning the amplitude of the \textit{alternating} Josephson current
for any frequency $\delta_\downarrow + \delta_\uparrow\ne 0$. 

\section{Spin-asymmetric Josephson effect in superlattices}
Experimental observations that may be described by the dc tunability
have been reported~\cite{Robinson2010}, while the experimental potential of the 
spin-asymmetric Josephson oscillations remains still untapped.
Here, we propose a setup to detect the spin-asymmetric Josephson effect and the tunable critical current
for instance in ultracold Fermi gases with existing experimental tools.

In ultracold atomic gases, the Josephson effect has been studied in experiments with Bose-Einstein condensates~\cite{internalbose,bosejose01,bosejose05,bosejose07}.
While there has been considerable progress in transport-type experiments
with ultracold fermions~\cite{Brantut2012,Stadler2012},
the Josephson effect remains still unobserved in Fermi gases.
However, the recent emergence of highly tunable optical lattice setups
of multi-spin systems~\cite{Mandel2003,Trotzky2008,Greif2013} 
offers an alternative way to approach the Josephson effect, and in particular,
its spin-asymmetric extension.

\begin{figure}
\centerline{\includegraphics[width=1.0\columnwidth]{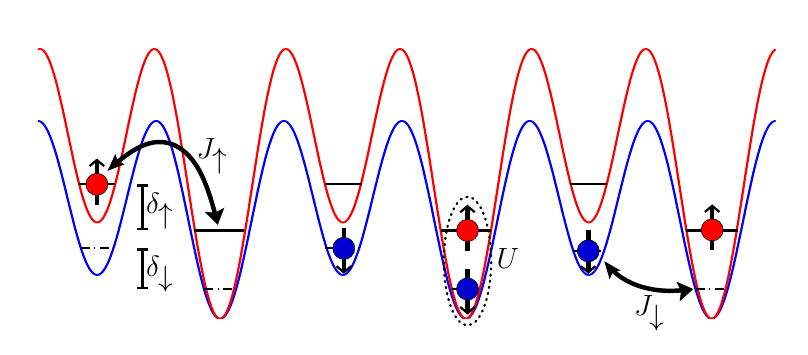}}
\caption{Spin-dependent superlattice setup to realize the
spin-asymmetric Josephson effect.
The $\uparrow$ spin (red ball) and $\downarrow$ spin (blue ball) 
tunnel between adjacent lattice sites with couplings 
$J_\uparrow$ and $J_\downarrow$, respectively.
The spin-dependent potential difference 
between neighboring sites is given by $\delta_\sigma$
and the on-site interaction strength by $U$. 
The observation of spin-asymmetric Josephson oscillations between 
adjacent lattice sites requires 
$\delta_\uparrow \neq \delta_\downarrow$, a condition which can
be met experimentally, e.g., in ultracold gases.}
\label{fig:system}
\end{figure}

We computationally predict that the spin-asymmetric Josephson oscillations take place
in a 1D spin-dependent superlattice 
where each pair of adjacent lattice sites is an analog of a Josephson junction. See Fig.~\ref{fig:system} for illustration.
For example in ultracold gases,
the superlattice can be constructed
by superimposing two optical lattices generated with
lasers of wavelengths $\lambda$ and $\lambda/2$.
There are several possibilities to obtain the required spin-dependence. 
First, one can utilize a Fermi-Fermi mixture of two different elements, 
such as $^6\mathrm{Li}$-$^{40}\mathrm{K}$ \cite{Taglieber2008,Willie2008} 
or $^6\mathrm{Li}$-Yb \cite{Hansen2013}. For instance, in the case 
of $^6\mathrm{Li}$-$^{40}\mathrm{K}$ the experimentally most suitable combination of hyperfine states 
would be $\ket{\uparrow}=\ket{1/2,1/2}_\mathrm{Li}$ 
and $\ket{\downarrow}=\ket{9/2,9/2}_\mathrm{K}$ \cite{Tiecke2010}. 
The spin-dependent Hubbard model parametrization arises from the different 
masses and the different optical properties of these elements. Secondly, 
recent theoretical proposals suggest that it is possible to create state-dependent 
lattice potentials for alkaline-earth and alkaline-earth-like atoms 
(e.g., Yb \cite{Fukuhara2007}, Sr \cite{DeSalvo2010,Tey2010}, and Dy \cite{Lu2012}) 
in two different internal states \cite{Daley2008,Daley2011}.
Note that the experimental setup can also have higher dimensionality for example 
if a 1D array of two-dimensional disks is used,
with each disk corresponding to one lattice site of the 1D model of this work. 

The system of Fig.~\ref{fig:system} is described by the Fermi--Hubbard Hamiltonian 
\begin{align}\label{fhsl2}
\hat{H}=&
-\sum_{i,\sigma} J_{\sigma} (\hat{c}^{\dagger}_{i+1,\sigma}\hat{c}_{i,\sigma}
+\mathrm{H.c.})
+U\sum_{i} \hat{n}_{i,\uparrow}\hat{n}_{i,\downarrow}\nonumber\\
&+\sum_{i,\sigma}\frac{\delta_{\sigma}}{2}(\hat{n}_{2i-1,\sigma}-\hat{n}_{2i,\sigma}).
\end{align} 
Here, $\hat{c}_{i,\sigma}$ ($\hat{c}^{\dagger}_{i,\sigma}$) is the fermionic annihilation 
(creation) operator for pseudo-spin $\sigma=\uparrow,\downarrow$ and lattice site $i$, and $\hat{n}_{i,\sigma}=\hat{c}^{\dagger}_{i,\sigma}\hat{c}_{i,\sigma}$ is the number operator. 
The nearest-neighbor tunneling matrix element, on-site interaction, and 
spin-dependent potential difference between adjacent lattice sites are 
denoted by $J_\sigma$, $U$, and $\delta_\sigma$, respectively.
The system is prepared in equilibrium at half-filling 
without a superlattice, i.e.\ $\delta_\uparrow=\delta_\downarrow=0$. At $t=0^+$, 
the spin-dependent superlattice potential is switched on,
resulting in $\delta_\uparrow \ne \delta_\downarrow$.

The crucial control parameter in the superlattice arrangement 
is the spin-dependent potential difference $\delta_\sigma$ between 
adjacent lattice sites. In the relevant region for the spin-asymmetric Josephson 
effect, $\delta_\sigma$ ranges from zero to roughly 10$J_\sigma$. 
In typical experiments with deep optical lattices, the depth of the 
lattice 
potential is about $10E_r$, where $E_r$ is the recoil energy. 
For such lattices the value of $J_\sigma$ is on the order of $0.01E_r$,
whereas the band gap is well above the recoil energy. 
Thus, the required values of $\delta_\sigma$ are more than an order of magnitude 
below the band gap and the depth of the lattice potential. As a result, the system is 
well described by the lowest-band Hubbard model of Eq.~\eqref{fhsl2} also in 
the presence of the spin-asymmetric potential.

\begin{figure} [ht!]
\includegraphics[width=\columnwidth]{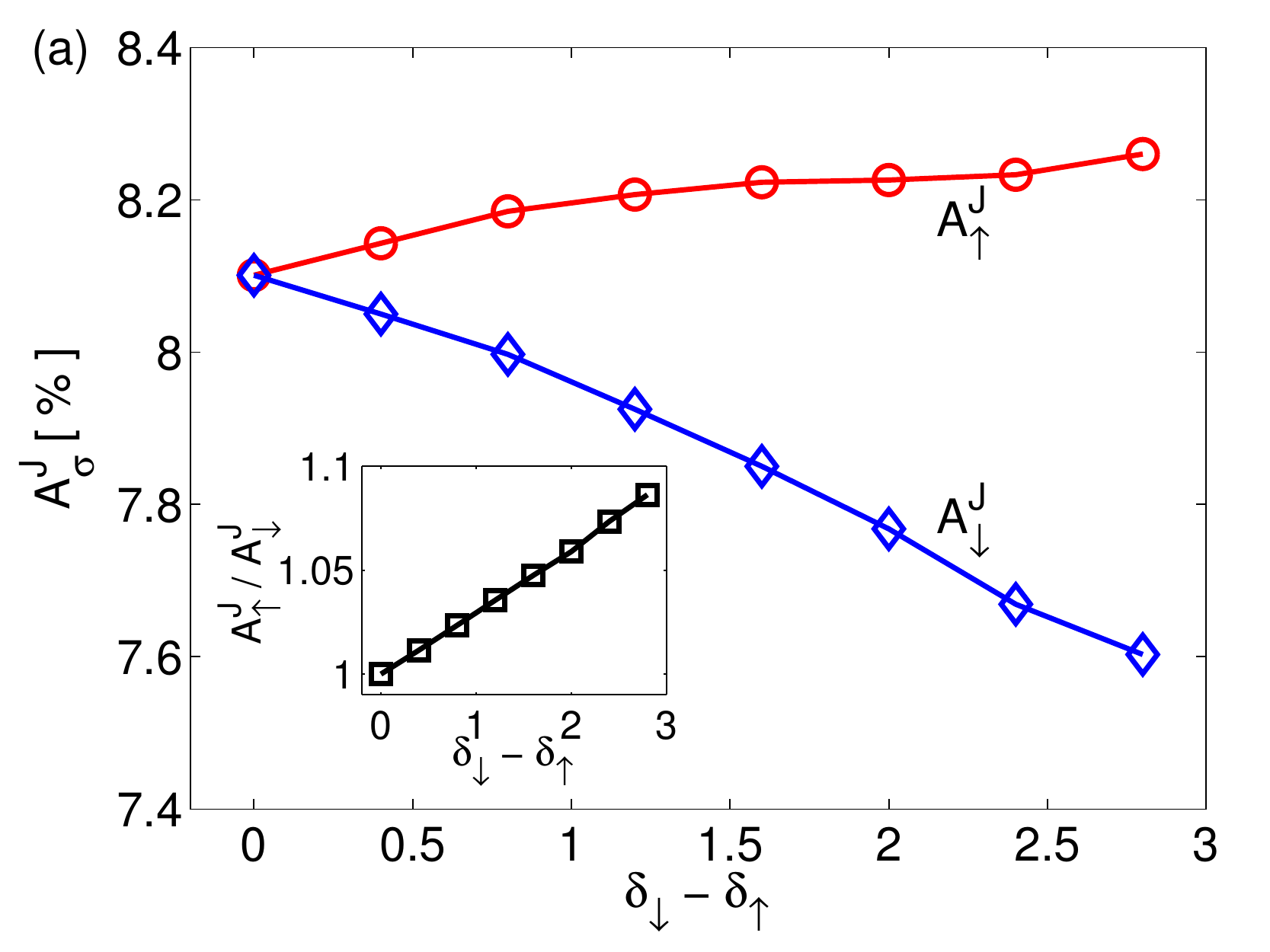}
\includegraphics[width=\columnwidth]{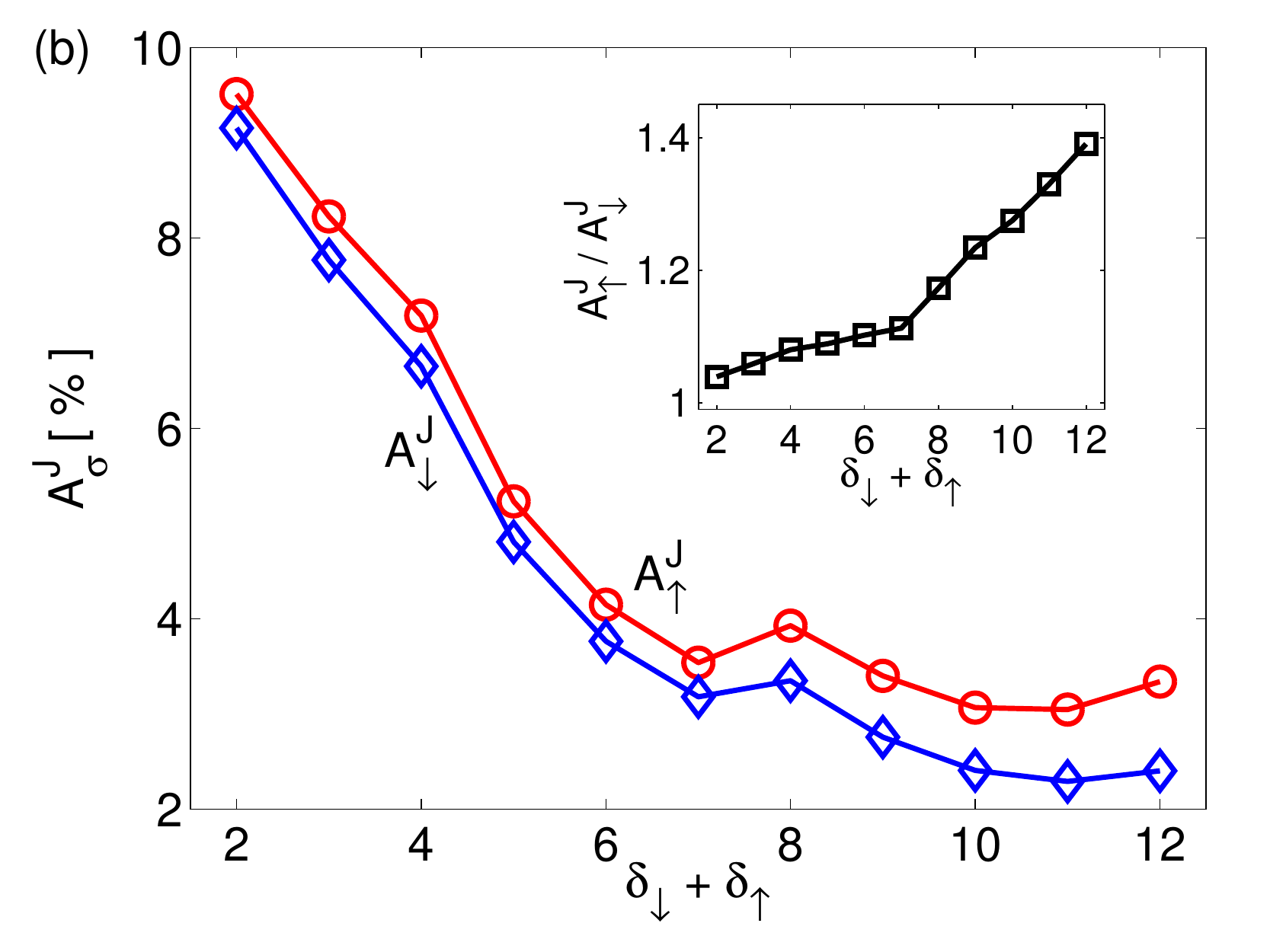}
\caption{Josephson oscillation amplitudes as a function of  $\delta_\downarrow-\delta_\uparrow$ and  $\delta_\downarrow+\delta_\uparrow$.
(a) The amplitudes of the Josephson oscillation, $A^J_\uparrow$ 
(red line with circles) and $A^J_\downarrow$
(blue line with diamonds) as a function of $\delta_\downarrow-\delta_\uparrow$,
when $\delta_\downarrow+\delta_\uparrow=3.0$,
given in proportion to the initial filling fraction of 0.5. 
For $\delta_\uparrow = \delta_\downarrow$, there is no difference 
in the amplitudes and the system displays the standard Josephson effect. 
For increasing $\delta_\downarrow-\delta_\uparrow$, the amplitudes deviate 
from the balanced value with $A^J_\uparrow$ increasing and $ A^J_\downarrow$ 
decreasing, and thus spin-asymmetric Josephson oscillations are observed. 
Inset: the relative asymmetry $A^J_\uparrow/A^J_\downarrow$. 
For greater values of $\delta_\downarrow-\delta_\uparrow$, 
an asymmetry of 9\% is obtained.
(b) The amplitude of the Josephson oscillation, $A^J_\sigma$,
as a function of $\delta_\downarrow+\delta_\uparrow$
with $\delta_\downarrow-\delta_\uparrow=2.0$. 
The greatest, hence more easily detectable, values of the amplitudes 
are obtained for small $\delta_\downarrow+\delta_\uparrow$. 
Inset: the relative asymmetry $A^J_\uparrow/A^J_\downarrow$.
Here, significant values of asymmetry up to 39\% are obtained.
In both (a) and (b), the amplitudes are so large that they may be
detected with existing imaging techniques.
}
\label{fig:result1}
\end{figure}


The dynamics of the superlattice system is simulated with 
the TEBD numerical method~\cite{vidal2003,daley2004,vidal2004}.
We study a system with $L=50$ lattice sites and matrix product state bond dimension
$\chi=150$. These parameters suffice to make finite size effects negligible and to restrain
the effect of any numerical artifacts on the Josephson oscillations.
For simplicity, we consider here the case $J_\uparrow=J_\downarrow$,
but our observations remain valid for $J_\uparrow\ne J_\downarrow$.
We set $\hbar=1$ and give all energies and frequencies in the units of 
$J_\uparrow$ ($J_\uparrow=1$). 
We focus on the attractive interaction $U=-10$ ($5<|U|<15$ would give similar results) with simulation time $t_{\mathrm{final}}=120$ to reach high accuracy in the Fourier transforms.

Our observable is the average particle number on odd lattice sites 
\begin{align}
N_{\sigma,\mathrm{odd}}(t)=\frac{1}{L_\mathrm{odd}}\sum_{i=1}^{L_\mathrm{odd}} N_{\sigma,2i-1}(t),
\end{align}
where $L_\mathrm{odd}=L/2$, since the particle number is the directly measurable quantity in an ultracold gas setup, as opposed to the current. 
We identify the Josephson oscillations between odd and even lattice sites
from the Fourier transformation $N_{\sigma,\mathrm{odd}}(\omega)$. 
The Josephson frequency is the same for both spin components even
in the presence of spin-asymmetric potentials.
There are also single-particle processes present, but the Josephson oscillations
dominate the physics.

In Fig.~\ref{fig:result1}(a) we exhibit how the Josephson oscillation amplitudes
of each spin component, $A^J_\uparrow$ and $A^J_\downarrow$, 
become unequal when spin-dependent potentials are applied i.e.\ $\delta_\downarrow\ne\delta_\uparrow$,
characteristic of the spin-asymmetric Josephson effect.
Moreover, we show in Fig.~\ref{fig:result1}(b) that for a 
fixed value of $\delta_\downarrow-\delta_\uparrow$ 
the Josephson amplitudes $A^J_\sigma$
grow with decreasing $\delta_\downarrow+\delta_\uparrow$. 
On the other hand, the difference in the oscillation amplitudes 
grows towards higher values of $\delta_\downarrow+\delta_\uparrow$.

Based on Fig.~\ref{fig:result1} we suggest that the Josephson amplitudes are 
large enough to be imaged with existing experimental techniques,
and can be tuned substantially by varying the values of $\delta_\uparrow$ and $\delta_\downarrow$.
Furthermore, the amplitudes exhibit significant spin-asymmetry:
in Fig.~\ref{fig:result1}(b) 
$A^J_\uparrow/A^J_\downarrow$
rises to a remarkable  
$39$\%. In our simulations, 
we have found that the amplitude asymmetry
can grow well beyond $100\%$ when $\delta_\downarrow-\delta_\uparrow$ is further increased. 

\begin{figure}
\includegraphics[width=\columnwidth]{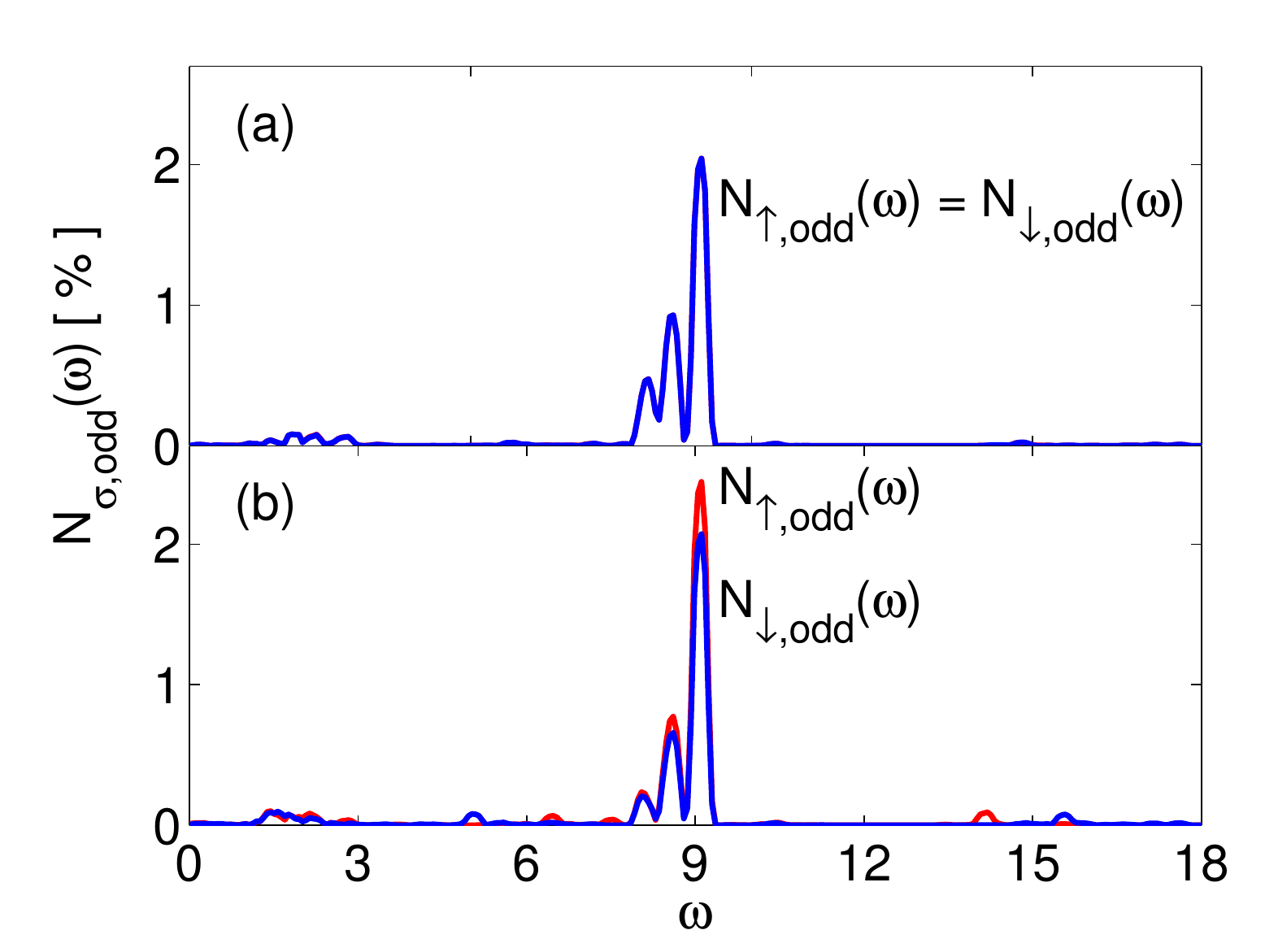}
\caption{Frequency spectrum of the particle number dynamics.
The Fourier transformation of the average particle number on
odd lattice sites, $N_{\sigma,\mathrm{odd}}(\omega)$, is given  
relative to the initial filling fraction of $0.5$.
Here, $\delta_\downarrow+\delta_\uparrow=9.0$, with 
(a) $\delta_\downarrow-\delta_\uparrow=0.0$
and
(b) $\delta_\downarrow-\delta_\uparrow=1.6$.
The peak structure about ${\omega=\delta_\downarrow+\delta_\uparrow=9.0}$ is the Josephson contribution. 
The Josephson signal is split into sub-peaks due 
to higher-order modifications to the Josephson frequency.
Note that the asymmetric potential leads to a clear difference in the Josephson amplitudes
of the spin components, while the Josephson frequency remains the same for both spins.
In addition to Josephson oscillations, there are minor contributions from single-particle 
and higher order processes at other frequencies.}
\label{fig:result2}
\end{figure}

We also find that in order to have experimentally observable particle number oscillations,
higher order modifications to the Josephson frequency 
emerge as demonstrated in Fig.~\ref{fig:result2}.
The Josephson peak is split into sub-peaks, and
the frequency difference between adjacent sub-peaks can be estimated from
perturbation theory to second order in $J_\sigma$, yielding
\begin{align}
\Delta \omega_J^{(2)}=2|U|\left(\frac{J_\uparrow^2}{U^2-\delta_\uparrow^2}+\frac{J_\downarrow^2}{U^2-\delta_\downarrow^2}\right)
\end{align}
For details, see Appendix~\ref{app:high}.
Moreover, the center of the Josephson peak structure is shifted away from the typical
Josephson frequency, $\delta_\downarrow+\delta_\uparrow$. The shift can be estimated as
\begin{align} 
\omega_{J,\mathrm{shift}}={\frac{2J_\uparrow^2\delta_\uparrow}{\delta_\uparrow^2-U^2}+\frac{2J_\downarrow^2\delta_\downarrow}{\delta_\downarrow^2-U^2}}.
\end{align}
An analogous shift can be found in the two-state problem, where the Rabi
frequency $\Omega_{\mathrm{Rabi}}$ is shifted from the detuning $\delta$ because of the finite coupling $J$. 
To second order in $J$, the Rabi frequency is
\begin{align} 
\Omega_{\mathrm{Rabi}}=\sqrt{\delta^2+J^2}\simeq \delta +\frac{J^2}{2\delta}.
\end{align}
The effect is also similar to the superexchange shift observed in Bose gases \cite{Trotzky2008}.
We emphasize that in spite of the higher order effects, the Josephson frequency remains the
same for both spin-components in the spin-asymmetric case. 

On a final note, we point out that in the dc limit the setup we propose has a fundamental 
difference to the theoretical model of a simple Josephson junction. First, the transport 
type arrangement where current is injected through the system is not straightforward in the 
superlattice. Second, the superfluid ground state has a zero phase difference between 
nearest-neighboring lattice sites. Therefore, the dc Josephson current 
would be zero in this superlattice arrangement.
Regarding the ac Josephson effect, the quantity $\delta_{\uparrow} + \delta_{\downarrow}$ 
is limited from below by the requirement of having a sufficient number of oscillations 
within the duration of the experiment, typically $100~\textrm{ms}$ scale.

\section{Summary}
In summary, our calculations suggest that the spin-asymmetric Josephson
effect can be realized in a spin-dependent superlattice, e.g., utilizing
a two-component ultracold Fermi gas. 
The amplitudes of the Josephson oscillations of each spin-component,
as well as their relative difference, are highly tunable, and can be detected
with state-of-the-art imaging techniques.
The spin-asymmetric Josephson effect reveals the existence of a single particle 
interference contribution in the standard Josephson current by splitting the
degeneracy of the intermediate states where this interference occurs. 
The dc limit of the spin-asymmetric Josephson effect provides an interesting 
viewpoint to the physics of the tunable critical current in SFIFS junctions. 
The observation of the asymmetric effect would extend the fundamental understanding of 
Josephson phenomena and the related high tunability of the critical current 
promises versatile Josephson devices.

\begin{acknowledgements}
We thank J. Kajala for useful discussions.
This work was supported by the Academy of Finland through 
its Centers of Excellence Programme (2012-2017) 
and under Projects No. 141039, No. 135000,
No. 251748 and No. 263347.
M.O.J.H. acknowledges financial support from the 
Finnish Doctoral Programme in Computational Sciences FICS.
F.M. acknowledges financial support from ERC Advanced Grant MPOES.
Computing resources were provided by CSC--the Finnish IT 
Centre for Science and the Aalto Science-IT Project.
\end{acknowledgements}

\appendix

\section{Connection between the tunable critical current in SFIFS junctions and the spin-asymmetric Josephson effect}\label{app:connection}

\setcounter{equation}{0}

Here, we consider in detail the connection between the spin-asymmetric 
Josephson effect~\cite{sorin,Heikkinen2010}
and the tunable critical current in SFIFS junctions~\cite{Bergeret2001}.
(Here S stands for superconductor, F for ferromagnet, and I for insulator.)
We demonstrate that at zero temperature these two scenarios are in fact equivalent,
provided that the exchange field of the ferromagnetic layers is not strong enough to
destroy the superconducting state.

Conceptually, the spin-asymmetric Josephson effect involves two superfluids or superconductors
connected by a tunneling coupling, see Figs.~\ref{fig:basics1} and \ref{fig:basics2}.
Initially, the system is at equilibrium without any spin-dependent potentials,
and at $t=0^+$ such potentials are switched on, resulting in the spin-asymmetric
Josephson effect. In the context of the tunable critical
current in SFIFS junctions~\cite{Bergeret2001}, 
the magnetization of the ferromagnetic layers is present in the ground state of the system,
and the SF bilayer is considered a uniformly magnetized superconductor with an effective
exchange field. This assumption is valid if
the thickness of the S layer is below the superconducting coherence length, and the 
thickness of the F layer is below the condensate penetration length to the ferromagnet~\cite{Bergeret2001}.
Apart from affecting also the initial state of the system, this uniform effective exchange field
plays a role similar to the difference of the potentials for each spin-component
in the spin-asymmetric Josephson effect.

In the following, we consider both the effective exchange fields of the SFIFS junction
and the spin-asymmetric potentials within the same formalism.
We show that the exchange fields and the spin-asymmetric potentials have the
same contribution to the critical Josephson current, while the Josephson frequency
is a function of the spin-asymmetric potentials only.
We use the subindices
$L$ and $R$ to denote the left and right sides of the Josephson junction.
The Hamiltonian for the ground state of the system 
is $\hat{H}_0=\hat{H}_L+\hat{H}_R$ where $\hat{H}_L$ and $\hat{H}_R$
are the Hamiltonians of the left and right superconductors 
in the presence of an effective exchange field. 
To simplify the derivation, we assume
that without the exchange field, the left and right superconductors are identical,
and set $\hbar=1$.
The Hamiltonian $\hat{H}_L$ is
\begin{align}
\hat{H}_{L}=&
\sum_{k,\sigma} \xi_{k} \hat{n}_{L,k,\sigma}
+g\sum_{k} \hat{n}_{L,k,\uparrow}\hat{n}_{L,-k,\downarrow} \nonumber \\
&-h_{L}\sum_{k}(\hat{n}_{L,k,\uparrow}
-\hat{n}_{L,k,\downarrow}),
\end{align}
and the expression for $\hat{H}_R$ is similar. 
The operators $\hat{c}_{L,k,\sigma}$ and $\hat{c}^{\dagger}_{L,k,\sigma}$ 
are the fermionic annihilation 
and creation operators with momentum $k$ and pseudo-spin $\sigma=\uparrow,\downarrow$, and
$\hat{n}_{L,k,\sigma}=\hat{c}^{\dagger}_{L,k,\sigma}\hat{c}_{L,k,\sigma}$ is the number operator.
The kinetic energy of momentum state $k$ 
relative to the chemical potential is denoted by $\xi_k$ and the interaction strength and
effective exchange field are $g$ and $h_L$, respectively.
The tunneling Hamiltonian which is switched on at $t=0^+$ reads
\begin{align}
\hat{H}_{\Omega}=&\sum_{k,p,\sigma} \Omega_{k,p} (\hat{c}^{\dagger}_{L,k,\sigma}\hat{c}_{R,p,\sigma}+\mathrm{H.c.}) \nonumber \\
&+\sum_{k,\sigma}\frac{\delta_{\sigma}}{2}(\hat{n}_{L,k,\sigma}-\hat{n}_{R,k,\sigma}).
\end{align}
Here, $\delta_\sigma$ is the spin-dependent potential across the junction,
and $\Omega_{k,p}$ is the tunneling matrix element which couples
the left and right sides of the junction. 
For brevity, we include both the exchange field of the ferromagnet 
and the spin-asymmetric potential formally in the same 
total Hamiltonian $\hat{H}=\hat{H}_0+\hat{H}_\Omega$.
We present the most important part of the derivation separately for each
of the two scenarios.

Assuming uniform mean-field BCS pairing, the ground state Hamiltonian is diagonalized following
the standard BCS derivation. For computing the Josephson current, the expression
for the anomalous Green's function (in Matsubara formalism) is required.
The anomalous Green's function $F_L(k,\tau)=-\aver{T_\tau c_{L,k,\uparrow}(\tau)c_{L,-k,\downarrow}(0)}$, 
where $T_\tau$ is the time-ordering operator and the
angle brackets denote the thermodynamic average, for 
the ground state of the left superconductor is given by
\begin{align}
F_L(k,i\omega)
=u_kv_k\left( \frac{1}{i\omega-E^+_L(\xi_k)} - \frac{1}{i\omega+E^-_L(\xi_k)} \right),
\end{align}
with the notations
\begin{align}
E^\pm_L&=\mp h_L+\sqrt{\xi_k^2+\Delta^2},\nonumber\\
u_k&=\sqrt{\frac{1}{2}+\frac{\xi_k}{2\sqrt{\xi_k^2+\Delta^2}}},\nonumber\\
v_k&=\sqrt{\frac{1}{2}-\frac{\xi_k}{2\sqrt{\xi_k^2+\Delta^2}}}.
\end{align}
Here, $i\omega$ is the fermionic Matsubara frequency and $\Delta$ the BCS order parameter.
The expression for the Green's function for the right side of the junction is again similar.
Notice that at zero temperature the only dependence of the order parameter $\Delta$
on the exchange field is that $\Delta(h_{L/R})=0$ when $h_{L/R}>\Delta(h_{L/R}=0)$. 
Moreover, for $F_L^\dagger(k,\tau)=-\aver{T_\tau c^{\dagger}_{L,-k,\downarrow}(\tau)c^{\dagger}_{L,k,\uparrow}(0)}$ 
we have
\begin{align}
F^{\dagger}_L(k,i\omega)=[F_{L}(k,-i\omega)]^*=F_{L}(k,i\omega).
\end{align}
In the latter equality we have assumed a real gap.

The linear response derivation of the spin-asymmetric 
Josephson effect~\cite{Heikkinen2010} with respect to $\hat{H}_\Omega$
can be followed also in the presence of magnetization
up to the following expression for the Josephson current.
Taking the $\uparrow$ spin-component as the example, we have
\begin{align}
I_\uparrow^J(t)&=I_\uparrow^C(-\delta_\downarrow)\sin [(\delta_\uparrow+\delta_\downarrow)t+\varphi].
\end{align}
Here, $\varphi$ is the initial phase difference across the junction, 
while the critical current is
\begin{align}
I_\uparrow^C(-\delta_\downarrow)&=2\sum_{kp}\Omega_{kp}\Omega_{-k,-p} |\Pi_F(k,p,-\delta_\downarrow)|,
\end{align}
with 
\begin{align}
\Pi_F(k,p,i\omega)=\frac{1}{\beta}\sum_{i\chi}F_L(k,i\chi+i\omega)F^{\dagger}_R(p,i\chi),
\end{align}
where $\beta$ is the inverse temperature.

Let us first consider the standard SFIFS junction without the spin-asymmetric potentials, 
i.e. $\delta_\uparrow=\delta_\downarrow=0$, in which case also the Josephson frequency is zero.
Inserting the anomalous Green's function to the expression above, we find
by using standard Matsubara summation techniques
\begin{widetext}
\begin{align}
\Pi_F(k,p,i\omega)
=&\frac{1}{\beta}\sum_{i\chi} u_k v_ku_p v_p
\left( \frac{1}{i\omega+i\chi-E_L^+(k)}-\frac{1}{i\omega+i\chi+E_L^-(k)} \right)
\left( \frac{1}{i\chi-E_R^+(p)}-\frac{1}{i\chi+E_R^-(p)} \right)
\nonumber\\
=&\frac{1}{\beta} \sum_{i\chi} u_k v_k u_p v_p \left(
 \frac{1}{i\omega+i\chi-E_L^+(k)}\frac{1}{i\chi-E_R^+(p)}
- \frac{1}{i\omega+i\chi-E_L^+(k)}\frac{1}{i\chi+E_R^-(p)}\right.\nonumber\\
&\left.- \frac{1}{i\omega+i\chi+E_L^-(k)}\frac{1}{i\chi-E_R^+(p)}
+ \frac{1}{i\omega+i\chi+E_L^-(k)}\frac{1}{i\chi+E_R^-(p)}\right)
\nonumber\\
=& u_k v_ku_p v_p
\left(
\frac{n_F(E_L^+(k))-n_F(E_R^+(p))}{E_L^+(k)-E_R^+(p)-i\omega}
-\frac{n_F(E_L^+(k))-n_F(-E_R^-(p))}{E_L^+(k)+ E_R^-(p)-i\omega}\right.\nonumber\\
&\left. -\frac{n_F(-E_L^+(k))-n_F(E_R^-(p))}{-E_L^-(k)- E_R^+(p)-i\omega}
+\frac{n_F(-E_L^+(k))-n_F(-E_R^-(p))}{-E_L^-(k)+ E_R^-(p)-i\omega}\right).
\end{align}
\end{widetext}
Taking the limit $T=0$, where the Fermi function is $n_F(E<0)=1$ and $n_F(E>0)=0$, we obtain
\begin{align}
\Pi_F(k,p,i\omega)
=
&\frac{u_k v_ku_p v_p}{E_L^+(k)+ E_R^-(p)-i\omega}\nonumber \\
&+\frac{u_k v_ku_p v_p}{E_L^-(k)+ E_R^+(p)+i\omega}.
\end{align}
Using the notation $E_0(k)=\sqrt{\xi_k^2+\Delta^2}$ and inserting 
the quasi-particle energies we have\medskip 
\begin{align}
\Pi_F(k,p,i\omega)
=
&\frac{u_k v_ku_p v_p}{E_0(k)+E_0(p)+h_R-h_L-i\omega}\nonumber \\
&+\frac{u_k v_ku_p v_p}{E_0(k)+E_0(p)-h_R+h_L+i\omega}.
\end{align}
At this point, we carry out the analytical continuation 
from Matsubara frequencies to real frequencies (here to $\omega=0$)
by taking $i\omega=i\eta^+$. We find the expression
\begin{widetext}
\begin{align}
\Pi_F(k,p,0)
=&
\frac{u_k v_ku_p v_p}{E_0(k)+E_0(p)+h_R-h_L-i\eta^+}
+\frac{u_k v_ku_p v_p}{E_0(k)+E_0(p)-(h_R-h_L)+i\eta^+}\nonumber\\
=&
u_k v_ku_p v_p
\left [\mathcal{P}\frac{1}{E_0(k)+E_0(p)+h_R-h_L}
+i\pi\delta\big(E_0(k)+E_0(p)+h_R-h_L\big)\right.\nonumber\\
&+\left.\mathcal{P}\frac{1}{E_0(k)+E_0(p)-(h_R-h_L)}-i\pi\delta \big(E_0(k)+E_0(p)-(h_R-h_L)\big) \right ].
\end{align}
\end{widetext}
Finally, when $h_R$ and $h_L$ are below the gap the $\delta$-function does not contribute
since $E_0(k)\ge\Delta$.
For the same reason, the principal value integral denoted with $\mathcal{P}$
becomes a regular one since the denominator does not 
contain any poles. We obtain
\begin{align}
I_\uparrow^C(0)=&2\sum_{kp}\Omega_{kp}\Omega_{-k,-p}u_k v_ku_p v_p \nonumber \\
&\times\Bigg [ \frac{1}{E_0(k)+E_0(p)+h_R-h_L}\nonumber \\ &+\frac{1}{E_0(k)+E_0(p)-(h_R-h_L)} \Bigg ].
\end{align}
The final part of the calculation would then involve carrying out the integration over the momenta
recovering the earlier result~\cite{Bergeret2001}.
However, since this integral has to be solved numerically (or analytically in an approximate form),
it is better to take the expression above as the point of comparison.

We now turn to the spin-asymmetric Josephson effect, i.e. have finite $\delta_\sigma$ 
present in the time-evolution and set $h_L=h_R=0$. The calculation of the critical current follows
the previous case with two differences. First, 
we now have $E^+_L(k)=E^-_L(k)=E_0(k)$ and $E^+_R(k)=E^-_R(k)=E_0(k)$.
Secondly, the analytical continuation for $I_\uparrow^C$ is calculated at $-\delta_\downarrow$
taking $i\omega=-\delta_\downarrow+i\eta^+$ and $\eta^+\rightarrow 0$. 
Assuming that the potentials $\delta_\sigma$ are below the gap $\Delta$
(this is the same parameter range as with the exchange fields above), 
we then derive the expression
\begin{align}
&I_\uparrow^C (-\delta_\downarrow)=2\sum_{kp}\Omega_{kp}\Omega_{-k,-p} u_k v_k u_p v_p \nonumber \\
&\times \Bigg [ \frac{1}{E_0(k)+E_0(p)+\delta_\downarrow}+\frac{1}{E_0(k)+E_0(p)-\delta_\downarrow}\Bigg ].
\end{align}
Notice that the dc limit of the spin-asymmetric Josephson effect 
corresponds to the condition 
$\delta_\downarrow=-\delta_\uparrow$, and not only when both of the potentials are zero.

At this point we may conclude that the critical current 
takes the same form both in the case of the magnetically tuned SFIFS junction
as well as the spin-asymmetric Josephson effect.
In the former case, the critical current is tuned by the difference of the effective
exchange fields, while in the latter case the potentials $\delta_\sigma$ (in the dc limit
with the constraint $\delta_\downarrow=-\delta_\uparrow$) act as the control parameters. 
In other words, the magnetization difference and the spin-asymmetric potentials 
are interchangeable at $T=0$.

The result above shows that the tunability of the dc Josephson current in SFIFS 
junctions can be explained in terms of the spin-asymmetric Josephson effect,
as discussed in the main text.
In particular, the four-state model of a single Cooper pair in a superposition
across the junction can be used to explain the origin of the tunability, see Ref.~\cite{Heikkinen2010}
for a more detailed description. 
In the dc limit, the Josephson frequency is constant (and equal to zero) while the energy
of the intermediate states of the tunneling process
can still be tuned relative to the paired states by controlling
$h_R-h_L$ (or $\delta_\downarrow-\delta_\uparrow$).

\section{\label{4state} Four-state model of the spin-asymmetric 
Josephson effect with general tunneling couplings}

In our previous work~\cite{Heikkinen2010}, the four-state model was formulated explicitly for a momentum
conserving tunneling matrix element, while in the context of a tunneling junction
as in the calculation above, the
tunneling matrix is not diagonal in momentum space. 
In the following, we show that the Josephson dynamics can be reduced to the four-state
description for a general form of the tunneling coupling.
More specifically, we show that in the second order of perturbation theory
the four-state description is valid i.e.\ the total current
can be given as a sum over all possible four state systems.

In the following calculation, 
we simplify the notation by replacing the 
$\uparrow,\downarrow$ and $L,R$ indices with one
generic spin-label $\sigma=1,2,3,4$, with 
$1=(\uparrow,L)$, $2=(\downarrow,L)$,
$3=(\uparrow,R)$, and $4=(\downarrow,R)$
and absorb the single particle potentials
to the variable $\xi_{k\sigma}$.
Thus, it is assumed that in the ground state $\ket{\psi(t=0)}$
there is (zero temperature) BCS pairing between states $1,2$, and between $3,4$.
Similarly, tunneling is assumed between states $1,3$ and between $2,4$.
In this notation, the Hamiltonian of the system is
\begin{align}
\hat{H}_0&=\sum_{k\sigma}\xi_{k\sigma}\hat{n}_{k\sigma}
+g\sum_{k}\hat{n}_{k,1}\hat{n}_{-k,2} 
+g\sum_{k}\hat{n}_{k,3}\hat{n}_{-k,4}, \nonumber\\
\hat{H}_\Omega &=\sum_{k,k'}\Omega_{k1,k'3}\hat{c}^{\dagger}_{k1}\hat{c}^{}_{k'3}+\mathrm{H.c.} 
+\sum_{k,k'}\Omega_{k2,k'4}\hat{c}^{\dagger}_{k2}\hat{c}^{}_{k'4}+\mathrm{H.c.} 
\nonumber \\ &=\sum_{k\sigma,k'\sigma'}\Omega_{k\sigma,k'\sigma'}
\hat{c}^{\dagger}_{k\sigma}\hat{c}^{}_{k'\sigma'}.
\end{align}
Notice that $\Omega_{k\sigma,k'\sigma'}^*=\Omega_{k'\sigma',k\sigma}$.

Again, the Josephson effect is derived in second order perturbation theory with
respect to $\hat{H}_\Omega(t)=\exp(i\hat{H}_0t)\hat{H}_\Omega\exp(-i\hat{H}_0t)$. 
Using the interaction picture,
the time evolution of the initial state $\ket{\psi(t=0)}$ to second order in the tunneling is
\begin{align} 
\ket{\psi(t)}= &\ket{\psi(0)} -i\int\limits_0^tdt_1\,\hat{H}_\Omega(t_1)\ket{\psi(0)} \nonumber \\
&-\int\limits_0^t\int\limits_0^{t_1}dt_1\,dt_2\,\hat{H}_\Omega(t_1)\hat{H}_\Omega(t_2)\ket{\psi(0)}.
\end{align}
The zeroth and first order contributions to the total particle number of state $\nu$ defined by
$\aver{\hat{n}_\nu(t)}=\bra{\psi(t)}\hat{n}_\nu(t)\ket{\psi(t)}$ are the initial particle number and zero, respectively.
(Here, the particle number is a slightly more convenient quantity to calculate, 
and the current is given by its time derivative.) 
The second order contribution to $\aver{\hat{n}_\nu(t)}$ is
\begin{widetext}
\begin{align}
\aver{\hat{n}_\nu(t)}_2
=&\bra{\psi(0)}  \int\limits_0^tdt_2\,\hat{H}_\Omega(t_2) \hat{n}_\nu  \int\limits_0^tdt_1\,\hat{H}_\Omega(t_1) \ket{\psi(0)}-\bra{\psi(0)}  \int\limits_0^t\int\limits_0^{t_1}dt_1\,dt_2\, 
\hat{n}_\nu  \hat{H}_\Omega(t_2)\hat{H}_\Omega(t_1) \ket{\psi(0)}+\mathrm{H.c.}\nonumber\\
=
&\int\limits_0^t\int\limits_0^{t}dt_1\,dt_2\, \sum_{k\sigma,k'\sigma'}\sum_{p\mu,p'\mu'}\Omega_{k\sigma,k'\sigma'}\Omega_{p\mu,p'\mu'}\nonumber\\
&\times\bra{\psi(0)}
\exp(i\hat{H}_0t_2) \hat{c}^{\dagger}_{k\sigma}\hat{c}^{}_{k'\sigma'} \exp(-i\hat{H}_0t_2)
\hat{n}_\nu
\exp(i\hat{H}_0t_1) \hat{c}^{\dagger}_{p\mu}\hat{c}^{}_{p'\mu'} \exp(-i\hat{H}_0t_1)
\ket{\psi(0)}\nonumber\\
&-\int\limits_0^t\int\limits_0^{t_1}dt_1\,dt_2\, \sum_{k\sigma,k'\sigma'}\sum_{p\mu,p'\mu'}\Omega_{k\sigma,k'\sigma'}\Omega_{p\mu,p'\mu'}\nonumber\\
&\times
\bra{\psi(0)}
\hat{n}_\nu
\exp(i\hat{H}_0t_2) \hat{c}^{\dagger}_{k\sigma}\hat{c}^{}_{k'\sigma'} \exp(-i\hat{H}_0t_2)
\exp(i\hat{H}_0t_1) \hat{c}^{\dagger}_{p\mu}\hat{c}^{}_{p'\mu'} \exp(-i\hat{H}_0t_1)
\ket{\psi(0)}+\mathrm{H.c.}
\label{simplifythis}
\end{align}

\end{widetext}

In the first term of the last expression above, $k\sigma$ has been 
relabeled as $k'\sigma'$ and vice versa. 
We then simplify equation~\eqref{simplifythis}
by taking into account the specific form of the initial state.
Let us consider as an example the case $\mu=1$.
For $\mu=1$, the tunneling matrix element $\Omega_{p\mu,p'\mu'}$ sets directly $\mu'=3$.
For indices $(\sigma,\sigma')$ we have similarly the possibilities $(\sigma,\sigma')=(1,3),(3,1),(2,4),(4,2)$.
Now, the operators $\hat{n}_\nu$ and $\hat{H}_0$ above
do not change the number of unpaired particles, while 
the opposite is true for 
$\hat{c}^{\dagger}_{k\sigma}\hat{c}^{}_{k'\sigma'}$
and $\hat{c}^{\dagger}_{p1}\hat{c}^{}_{p'3}$.
For example the operator $\hat{c}^{\dagger}_{p1}\hat{c}^{}_{p'3}$ 
acting on a state with no unpaired particles
would create an unpaired fermion on states $\ket{p1}$ and $\ket{-p'4}$ (since 
the fermion on $\ket{p'3}$ is removed).
The remaining operator $\hat{c}^{\dagger}_{k\sigma}\hat{c}^{}_{k'\sigma'}$ then
has to act so that both of these unpaired states are removed (either by annihilating
the remaining fermion or by creating the missing fermion), or otherwise the resulting state
is orthogonal to $\bra{\psi(0)}$ and the matrix element is zero.

Since $\hat{n}_\nu$ and $\hat{H}_0$ are not relevant for finding the surviving matrix elements,
we leave these operators out of the notation in the following.
For the spin indices $(\sigma,\sigma')=(1,3),(4,2)$ we find
\begin{align}
\bra{\psi(0)}
\hat{c}^{\dagger}_{k1}\hat{c}^{}_{k'3}
\hat{c}^{\dagger}_{p1}\hat{c}^{}_{p'3}
\ket{\psi(0)} \equiv 0,\nonumber\\
\bra{\psi(0)}
\hat{c}^{\dagger}_{k4}\hat{c}^{}_{k'2} 
\hat{c}^{\dagger}_{p1}\hat{c}^{}_{p'3} 
\ket{\psi(0)} \equiv 0.
\end{align}
In other words, these arrangements of spin indices can only break Cooper pairs
and the matrix elements vanish.

For the indices $(\sigma,\sigma')=(3,1),(2,4)$ we find non-zero matrix-elements
for particular momenta $(k,k')$. 
Again, taking into account the Cooper pairing of $\ket{\psi(t=0)}$
we find these matrix elements to be
\begin{align}
\bra{\psi(0)}
\hat{c}^{\dagger}_{k3}\hat{c}^{}_{k'1}
\hat{c}^{\dagger}_{p1}\hat{c}^{}_{p'3}
\ket{\psi(0)} &=
\delta_{k,p'}\delta_{k',p}\nonumber \\
&\times \bra{\psi(0)}
\hat{c}^{\dagger}_{k3}\hat{c}^{}_{k'1}
\hat{c}^{\dagger}_{k'1}\hat{c}^{}_{k3}
\ket{\psi(0)},\nonumber\\
\bra{\psi(0)}
\hat{c}^{\dagger}_{k2}\hat{c}^{}_{k'4} 
\hat{c}^{\dagger}_{p1}\hat{c}^{}_{p'3} 
\ket{\psi(0)} &=
\delta_{k,-p}\delta_{k',-p'}\nonumber \\
&\times
\bra{\psi(0)}
\hat{c}^{\dagger}_{k2}\hat{c}^{}_{k'4}
\hat{c}^{\dagger}_{-k1}\hat{c}^{}_{-k'3}
\ket{\psi(0)}.
\end{align}

Returning to the full notation, the second order result of Eq.~\eqref{simplifythis} is then written as
\begin{widetext}
\begin{align}
\aver{\hat{n}_\nu(t)}_2=
&\int\limits_0^t\int\limits_0^{t}dt_1\,dt_2\, \sum_{k\sigma,k'\sigma'}\Omega_{k\sigma,k'\sigma'}\Omega_{k'\sigma',k\sigma}\nonumber\\
&\times\bra{\psi(0)}
\exp(i\hat{H}_0t_2) \hat{c}^{\dagger}_{k\sigma}\hat{c}^{}_{k'\sigma'} \exp(-i\hat{H}_0t_2)
\hat{n}_\nu
\exp(i\hat{H}_0t_1) \hat{c}^{\dagger}_{k'\sigma'}\hat{c}^{}_{k\sigma} \exp(-i\hat{H}_0t_1)
\ket{\psi(0)}\nonumber\\
&-\int\limits_0^t\int\limits_0^{t_1}dt_1\,dt_2\, \sum_{k\sigma,k'\sigma'}\Omega_{k\sigma,k'\sigma'}\Omega_{-k\bar{\sigma},-k'\bar{\sigma}'}\nonumber\\
&\times\bra{\psi(0)}
\exp(i\hat{H}_0t_2) \hat{c}^{\dagger}_{k\sigma}\hat{c}^{}_{k'\sigma'} \exp(-i\hat{H}_0t_2)
\hat{n}_\nu
\exp(i\hat{H}_0t_1) \hat{c}^{\dagger}_{-k\bar{\sigma}}\hat{c}^{}_{-k'\bar{\sigma}} \exp(-i\hat{H}_0t_1)
\ket{\psi(0)}+\mathrm{H.c.}\nonumber\\
&+\int\limits_0^t\int\limits_0^{t}dt_1\,dt_2\, \sum_{k\sigma,k'\sigma'}\Omega_{k\sigma,k'\sigma'}\Omega_{k'\sigma',k\sigma}\nonumber\\
&\times
\bra{\psi(0)}
\hat{n}_\nu
\exp(i\hat{H}_0t_2) \hat{c}^{\dagger}_{k\sigma}\hat{c}^{}_{k'\sigma'} \exp(-i\hat{H}_0t_2)
\exp(i\hat{H}_0t_1) \hat{c}^{\dagger}_{k'\sigma'}\hat{c}^{}_{k\sigma} \exp(-i\hat{H}_0t_1)
\ket{\psi(0)}\nonumber\\
&-\int\limits_0^t\int\limits_0^{t_1}dt_1\,dt_2\, \sum_{k\sigma,k'\sigma'}\Omega_{k\sigma,k'\sigma'}\Omega_{-k\bar{\sigma},-k'\bar{\sigma}'}\nonumber\\
&\times
\bra{\psi(0)}
\hat{n}_\nu
\exp(i\hat{H}_0t_2) \hat{c}^{\dagger}_{k\sigma}\hat{c}^{}_{k'\sigma'} \exp(-i\hat{H}_0t_2)
\exp(i\hat{H}_0t_1) \hat{c}^{\dagger}_{-k\bar{\sigma}}\hat{c}^{}_{-k'\bar{\sigma}'} \exp(-i\hat{H}_0t_1)
\ket{\psi(0)}+\mathrm{H.c.}
\end{align}
\end{widetext}
Here we have used the notation $\bar{\sigma}$ for opposite spin, i.e.\ $\bar{1}=2$, $\bar{2}=1$,
$\bar{3}=4$, and $\bar{4}=3$.
Finally we take the summation over the momenta and spins in front of the whole expression.
The result is then of the form
\begin{align}
\aver{\hat{n}_\nu(t)}_2=&\sum_{k\sigma,k'\sigma'}[\mathrm{four~state~system~of~transitions~}\nonumber\\
&(k\sigma\leftrightarrow k'\sigma')
\mathrm{~and~}(-k\bar{\sigma}\leftrightarrow -k'\bar{\sigma}')],
\end{align}
i.e.\ a summation over all possible four state systems. A single term of this sum
is equivalent to the second order perturbation theory result with
only the two couplings 
$\Omega_{k\sigma,k'\sigma'}$
and $\Omega_{-k\bar{\sigma} -k'\bar{\sigma}'}$ with the relevant initial
state being the superposition of the Cooper pairs of states $(k\sigma,-k\bar{\sigma})$
and $(k'{\sigma}', -k'\bar{\sigma}')$. More precisely, this superposition originates
from the initial many-body state as follows. The full initial state is a combination
of two BCS states
\begin{align}
\ket{\psi(t=0)}=
\prod_{p} (u_p+v_p \hat{c}^\dagger_{p1}\hat{c}^\dagger_{-p2})
\prod_{p'} (u_{p'}+v_{p'} \hat{c}^\dagger_{p'3}\hat{c}^\dagger_{-p'4}) \ket{\emptyset}.
\end{align}
For each pair of couplings $\Omega_{k1,k'3}$ and $\Omega_{-k2 -k'4}$ we
rewrite the initial state as
\begin{align}
\ket{\psi(t=0)}=&
(u_k+v_k \hat{c}^\dagger_{k1}\hat{c}^\dagger_{-k2})(u_{k'}+v_{k'} \hat{c}^\dagger_{k'3}\hat{c}^\dagger_{-k'4})\ket{\zeta},\nonumber\\
\ket{\zeta}=&\prod_{p \ne k} (u_p+v_p \hat{c}^\dagger_{p1}\hat{c}^\dagger_{-p2})\nonumber \\
&\times \prod_{p'\ne k'} (u_{p'}+v_{p'} \hat{c}^\dagger_{p'3}\hat{c}^\dagger_{-p'4}) \ket{\emptyset},
\end{align}
and further as
\begin{align}
\ket{\psi(t=0)}=&
\Big(u_ku_{k'}+v_ku_{k'}  \hat{c}^\dagger_{k1}\hat{c}^\dagger_{-k2}+u_{k} v_{k'} \hat{c}^\dagger_{k'3}\hat{c}^\dagger_{-k'4}\nonumber \\
&+v_kv_{k'}\hat{c}^\dagger_{k1}\hat{c}^\dagger_{-k2}\hat{c}^\dagger_{k'3}\hat{c}^\dagger_{-k'4}\Big)\ket{\zeta}.
\end{align}
From this form we see that the $u_ku_{k'}$-term corresponds to an empty state
with respect to $\Omega_{k1,k'3}$ and $\Omega_{-k2 -k'4}$, and does not contribute to the dynamics.
Similarly, the $v_kv_{k'}$-term does not contribute to the dynamics since it is Pauli blocked.
The remaining superposition of the two states $(k1,-k2)$ and $(k'3,-k'4)$ is then formally
the initial state of the time-evolution under $\Omega_{k1,k'3}$ and $\Omega_{-k2 -k'4}$.
The two intermediate states $(k1,-k'4)$ and $(-k2,k'3)$ are also required to describe
the time-evolution, and thus the four-state system is complete.

To conclude, the analysis of the (spin-asymmetric) Josephson effect 
presented in our previous work~\cite{Heikkinen2010} can indeed be applied to a general
tunneling coupling, and not just a momentum conserving one, which 
is precisely what we set out to prove.
Notice, however, that the derivation above holds only to the second order of perturbation theory.
In higher orders of the perturbation expansion there are terms 
present which, e.g., involve more than two momentum states.

\section{Higher-order corrections to Josephson oscillations}\label{app:high}

In Fig.~\ref{fig:result2} of the main text, we show that for experimentally observable particle oscillations the Josephson contribution consists of several sub-peaks instead of a single spin-asymmetric signal. Moreover, the center peak in Fig.~\ref{fig:result2} is slightly shifted from the typical Josephson frequency of $\delta_\uparrow+\delta_\downarrow$. Here, we elucidate the origin of these effects by studying smaller lattice systems with second-order time-independent perturbation theory with respect to the tunneling matrix elements in Eq.~\eqref{fhsl2} of the main text. 

The shift of the Josephson frequency from $\delta_\uparrow+\delta_\downarrow$ can be obtained by considering a half-filled Hubbard dimer (a two-site system with two particles) governed by the Fermi--Hubbard Hamiltonian in Eq.~\eqref{fhsl2} of the main text. Here, the unperturbed Hamiltonian is given by
\begin{align}
\hat{H}_0 = U \left(\hat{n}_{1,\uparrow}\hat{n}_{1,\downarrow}+\hat{n}_{2,\uparrow}\hat{n}_{2,\downarrow}\right)+\sum_{\sigma=\uparrow,\downarrow}\frac{\delta_\sigma}{2}\left(\hat{n}_{1,\sigma}-\hat{n}_{2,\sigma}\right),
\end{align}
and the perturbation is
\begin{align}\label{hopping}
\hat{H}'=- \sum_{\sigma=\uparrow,\downarrow} J_\sigma \left(\hat{c}^{\dagger}_{2,\sigma} \hat{c}_{1,\sigma}+\mathrm{H.c.}\right).
\end{align}

The Josephson oscillations (which are a second order effect in $J_\sigma$) occur between the paired states
\begin{align}
|1\rangle &= |\uparrow \downarrow, \emptyset \rangle, \\
|2\rangle &= |\emptyset,\uparrow \downarrow \rangle,
\end{align}
with unperturbed energies
\begin{align}
E_1^{(0)}&=U+\frac{\delta_\uparrow}{2}+\frac{\delta_\downarrow}{2},\\
E_2^{(0)}&=U-\frac{\delta_\uparrow}{2}-\frac{\delta_\downarrow}{2}.
\end{align}
The first order corrections to the energies, $E_n^{(1)}=\langle n^{(0)}|\hat{H}'|n^{(0)}\rangle$, are zero, while the second order corrections can be calculated from
\begin{align}\label{2ndcorr}
E_n^{(2)}=\sum_{m \neq n} \frac{\left| \langle m^{(0)}|\hat{H}'|n^{(0)}\rangle \right|^2}{E_n^{(0)}-E_m^{(0)}},
\end{align} 
yielding
\begin{align}
E_1^{(2)}&=\frac{J_\uparrow^2}{U+\delta_\uparrow}+\frac{J_\downarrow^2}{U+\delta_\downarrow},\\
E_2^{(2)}&=\frac{J_\uparrow^2}{U-\delta_\uparrow}+\frac{J_\downarrow^2}{U-\delta_\downarrow}.
\end{align}
Thus, we obtain the perturbed Josephson frequency as (here, $\hbar=1$)
\begin{align}
\omega_J^{(2)}&=\left| E_1^{(0)}+E_1^{(2)}-\left(E_2^{(0)}+E_2^{(2)} \right) \right|\nonumber \\
&=\delta_\uparrow+\delta_\downarrow+\frac{2J_\uparrow^2 \delta_\uparrow}{\delta_\uparrow^2-U^2}+\frac{2J_\downarrow^2 \delta_\downarrow}{\delta_\downarrow^2-U^2},
\end{align}
where the last two terms give the shift in the frequency.

To explain the emergence of several Josephson signals, it suffices to investigate a four-site system described by the Hamiltonian in Eq.~\eqref{fhsl2} of the main text. We again take the unperturbed Hamiltonian to be
\begin{align}
\hat{H}_0 = U \sum_{i=1}^4 \hat{n}_{i,\uparrow}\hat{n}_{i,\downarrow}+\sum_{\sigma=\uparrow,\downarrow}\frac{\delta_\sigma}{2}\left(\hat{n}_{1,\sigma}-\hat{n}_{2,\sigma}+\hat{n}_{3,\sigma}-\hat{n}_{4,\sigma} \right),
\end{align}
while the perturbation reads
\begin{align}\label{hopping4}
\hat{H}'=-\sum_{i=1}^3 \sum_{\sigma=\uparrow,\downarrow} J_\sigma \left(\hat{c}^{\dagger}_{i+1,\sigma} \hat{c}_{i,\sigma}+\mathrm{H.c.}\right).
\end{align}

As an example, we consider the following two paired states (other states are treated similarly)
\begin{align}
|1\rangle &= |\emptyset, \uparrow \downarrow, \emptyset, \uparrow \downarrow \rangle,\\
|2\rangle &=|\uparrow \downarrow, \emptyset, \uparrow \downarrow, \emptyset \rangle,
\end{align}
which Josephson oscillate  with the state
\begin{align}
|3\rangle &= \frac{1}{\sqrt{2}}\left( |\uparrow \downarrow, \emptyset, \emptyset, \uparrow \downarrow \rangle + | \emptyset, \uparrow \downarrow, \uparrow \downarrow, \emptyset \rangle \right).
\end{align}
The unperturbed energies of the above states are given by $E_1^{(0)}=2U-\delta_\uparrow-\delta_\downarrow$, $E_2^{(0)}=2U+\delta_\uparrow+\delta_\downarrow$, and $E_3^{(0)}=2U$, respectively. We can then calculate the frequency of the unperturbed Josephson oscillations as $\omega_J^{(0)}=\left| E_1^{(0)}-E_3^{(0)} \right|=\left| E_2^{(0)}-E_3^{(0)} \right|=\delta_\uparrow+\delta_\downarrow$ which is precisely the typical Josephson frequency. 

The second order corrections can be obtained from Eq.~\eqref{2ndcorr} ($E_n^{(1)}=0$ as above). We obtain
\begin{align}
E_1^{(2)}&=\frac{3J_\uparrow^2}{U-\delta_\uparrow}+\frac{3J_\downarrow^2}{U-\delta_\downarrow},\\
E_2^{(2)}&=\frac{3J_\uparrow^2}{U+\delta_\uparrow}+\frac{3J_\downarrow^2}{U+\delta_\downarrow},\\
E_3^{(2)}&=2U\left( \frac{J_\uparrow^2}{U^2-\delta_\uparrow^2}+ \frac{J_\downarrow^2}{U^2-\delta_\downarrow^2}\right).
\end{align}

The perturbed Josephson frequencies are then given by
\begin{align}
\omega_{J,13}^{(2)}&=\left| E_1^{(0)}+E_1^{(2)}-\left(E_3^{(0)}+E_3^{(2)} \right) \right|\nonumber \\
&=\delta_\uparrow+\delta_\downarrow+\frac{2J_\downarrow^2}{\delta_\downarrow-U}+\frac{2J_\uparrow^2}{\delta_\uparrow-U}+\frac{J_\downarrow^2}{\delta_\downarrow+U}+\frac{J_\uparrow^2}{\delta_\uparrow+U},
\end{align}
and
\begin{align}
\omega_{J,23}^{(2)}&=\left| E_2^{(0)}+E_2^{(2)}-\left(E_3^{(0)}+E_3^{(2)} \right) \right|\nonumber \\
&=\delta_\uparrow+\delta_\downarrow+\frac{2J_\downarrow^2}{\delta_\downarrow+U}+\frac{2J_\uparrow^2}{\delta_\uparrow+U}+\frac{J_\downarrow^2}{\delta_\downarrow-U}+\frac{J_\uparrow^2}{\delta_\uparrow-U}.
\end{align}
We see that the Josephson frequencies of these processes are no longer equal and the difference reads
\begin{align}
\Delta \omega_J^{(2)}&=\omega_{J,13}^{(2)}-\omega_{J,23}^{(2)}\nonumber \\
&=2|U|\left(\frac{J_\uparrow^2}{U^2-\delta_\uparrow^2}+ \frac{J_\downarrow^2}{U^2-\delta_\downarrow^2}\right).
\end{align}
This is approximately the frequency separation between adjacent sub-peaks in Fig.~\ref{fig:result2} of the main text.

\end{document}